\documentclass{article}%
\usepackage{amsmath}
\usepackage{amsfonts}
\usepackage{amssymb}
\usepackage{graphicx}%
\providecommand{\U}[1]{\protect\rule{.1in}{.1in}}
\begin{document}

\title{An alternative to the gauge theoretic setting }
\author{Bert Schroer\\CBPF, Rua Dr. Xavier Sigaud 150 \\22290-180 Rio de Janeiro, Brazil\\and Institut fuer Theoretische Physik der FU Berlin, Germany}
\date{November 2010}
\maketitle

\begin{abstract}
The standard formulation of gauge theories results from the Lagrangian
(functional integral) quantization of classical gauge theories. A more
intrinsic qunantum theoretical access in the spirit of Wigner's representation
theory shows that there is a fundamental clash between the pointlike
localization of zero mass (vector, tensor) potentials and the Hilbert space
(positivity, unitarity) structure of QT. The quantization approach has no
other way than to stay with pointlike localization and sacrifice the Hilbert
space whereas the approach build on the intrinsic quantum concept of modular
localization keeps the Hilbert space and trades the conflict creating
pointlike generation with the tightest consistent localization:: semiinfinite
spacelike string localization. Whereas these potentials in the presence of
interactions stay quite close to associated pointlike field strength, the
interacting matter fields to which they are coupled bear the brunt of the
nonlocal aspect in that they are string.generated in a way which cannot be
undone by any differentiation.

The new stringlike approach to gauge theory also revives the idea of a
Schwinger-Higgs screening mechanism as a deeper and less metaphoric
description of the Higgs spontaneous symmetry breaking and its accompanying
tale about "God's particle" and its mass generation for all the other
particles... .

\end{abstract}

\section{Problematic aspects of the gauge theoretic setting of QED and QCD}

The standard formulation of gauge theories follows the quantization
parallelism to the classical Maxwell theory coupled to classical sources. A
object of significant classical computational importance is the
vectorpotential as it appears in the form of Lienard-Wiechert potentials of
classical charge distributions. It behaves like a covariant classical
vectorfield, apart from the fact that it cannot be uniquely fixed in terms of
Cauchy data. Except for this particularity it behaves as a classical covariant
causally propagating pointlike field.

Its quantized form entered the discourse of QFT almost from its beginnings,
and by 1929 the status of gauge theories has been competently expressed in a
review of the first phase of QFT \cite{Khar}\cite{Jor}. 

As we know nowadays, the imperfections were not that insignificant as the
author thought at that time, otherwise it would be difficult to understand
that it took another two decades to use indefinite metric Krein spaces in the
Gupta-Bleuler formalism\footnote{The application of the canonical quantization
rules to models involving vector-fields was always a point of contention and,
as will be shown here, the indefinite metric gauge theoretic formulation is by
no means the last word.} in order to make the quantum vectorpotentials
compatible with the structure of state space in QT; the nonabelian gauge
theories required a more complicated operator-ghost setting which was
developed much later and which bears the name of the initials BRST of the
protagonists Becchi, Rouet, Stora and Tuytin. These technical additions were
important for the renormalization project and the return to physical
observables, but they only affect some technical aspects of the quantization
transfer of the classical gauge concept and not the idea as such. In fact
apart from spacetime symmetries it is the only "symmetry" in classical
physics; the "inner" symmetries (flavour symmetries) are of pure QFT origin
and are a consequence of the superselection theory of localizable charges
\cite{Haag}. They are often red back into classical field theory with the
purpose to present a QFT model as a canonical or functional integral
quantization. 

In both cases one was forced to do the renormalization calculations involving
covariant pointlike fields in an indefinite metric setting; only at the end
one could expect to return to a physical Hilbert space via a GNS construction
based on the positive-semidefinite gauge invariant correlation functions. The
construction of non pointlike matter fields carrying a Maxwell- or Yang-Mills-
charge and their physical correlations, including those of matter fields,
remained an extremely difficult if not impossible task and.

It is not without irony that the post QED formalism for strong interactions
(the meson-nucleon interaction) of the 60s was conceptually much simpler. With
its mass gaps it represented precisely the arena in which all the
prerequisites for the derivation of the LSZ scattering theory including the
Haag-Ruelle derivation from the locality and spectral principles in a Wightman
setting hold; in fact it can hardly be called an accident that modern
scattering theory in QFT was developed at the time of strong interactions
between nucleons and mesons. 

The attempts to include QED into this ideal world of relativistic particle
physics (described by "interpolating" local quantum fields)
failed\footnote{The occurrence of infrared divergences in a model of QFT is
always an indication that there exist not yet understood spacetime
localization properties. Even though they can be patched up by momentum space
manipulations as the abandonment of scattering amplitudes in favor of certain
inclusive cross sections, there is, in the long run, no substitute for a
direct understanding in terms of spacetime localization properties.}. The
first difficulties were noted in form of perturbative infrared divergences in
scattering theory; they were bypassed by constructing quantities which are
apparently infrared convergent as the photon-inclusive cross sections for
scattering of charged particles \cite{B-N}\cite{YFS}. These calculations
received also nonperturbative support from the observation that the
K\"{a}ll\'{e}n Lehmann spectral functions of infraparticles which start with a
cut singularity at $p^{2}=m^{2}$ instead of a mass-shell delta function, in
that case the LSZ limits vanish. This is precisely what one expects of the
energy-momentum spectrum of indecomposable string-localized fields i.e.
string-localized fields which cannot be written as an integral along a string.
All these observations receive support from rigorous studies of the quantum
Gauss law \cite{Haag}.

Similar to the YFS infrared calculation \cite{YFS}, these cut singularities
which start at the mass shell are the result of summing of log singularities
which leads to a coupling dependent power cut. Since spectral functions are
positive measures, the strength of the cut must be smaller then that of the
delta function; the large p behavior is however not limited beyond the
temperedness restriction of the two-point distribution.

The reason why the construction of Gupta-Bleuler factor spaces, respectively
in case of BRST the construction of the cohomology space (related to
invariance under the nonlinear BRST transformation) leads to the unsolved
problems of charged matter is that precisely in this step the non-compact
localized physical charged operators have to emerge from their pointlike
unphysical gauge dependent pointlike counterparts. Maxwell quantum charge and
its particular type of semiinfinite stringlike generators are synonymous. In
the indefinite metric description the localization of gauge variant operators
had no physical meaning, it only offers the technical advantage of being able
to use the renormalization machinery of pointlike covariant fields. In fact in
using pointlike vectorpotentials one has to be extremely careful in order to
avoid incorrect conclusions about the absence of the quantum field theoretical
Aharonov-Bohm effect or the absence of Maxwell charges (see next section).

Instead of the quantization approach, which inevitably leads into the
well-known gauge theoretic formulation with all its physical problems and
technical attraction, we will set up an intrinsic framework which results from
combining Wigner's representation theoretical setting with the concept of
modular localization (next section), both having nothing in common with any
classical quantization parallelism. This leads automatically to
string-localized vectorpotentials acting in a Hilbert space with scale
dimension $d_{scale}=1$ which is the power counting prerequisite for the
existence of renormalizable interactions. Different from the pointlike gauge
theoretic formulation in the indefinite metric setting which is required by
quantization, the stringlike potential setting has the great advantage of
permitting an extension to interactions involving higher spin $s>1$ zero mass
fields: there always exist a tensor potential with $s_{scale}=1$ and a
coupling to matter fields which fulfills the power counting requirement at
least for certain couplings to low spin matter and to its own copies (viz
Yang-Mills models).

In this setting the notion of gauge has been replaced by the notion of
"causal, (modular\footnote{"Modular localization" stands for the intrinsic
(field-coordinatization-independent) formulation of causal relativistic
localization in QT. \ A formulation of QFT which highlights this property is
often called "local quantum physics" (LQP).}) localization". The results of
both approaches agree on local observables (always pointlike generated), on
that subalgebra there holds in fact the identity: local observables = gauge
invariant subalgebra. But on string-like objects (as matter fields carrying
Maxwell charges) which are genuinely noncompact string-localized, the gauge
formulation offers an unphysical formalism without a practicable way to
construct the important physical quantities which cannot be pointlike
generated. With the exception of selfcouplings between tensorpotentials
(Yang-Mills, selfcouplings between tensor potentials $g_{\mu\nu}(x,e)$ etc.),
the string-localization of the potentials remains quite
harmless\footnote{There are linear differential operators which reconvert the
potentials into pointlike field strengths. For the generation of an
irreducible set of operators the field strength are sufficient, but for
formulating the renormalizable interaction one needs the string-localized
potentials.}; but their mild nonlocal changes (moderated by the linear
connection to pointlike field strengths), affects the localization of matter
fields in a radical way which can not be undone by any linear operation.
Whereas the return of stringlike potentials to pointlike fields is quite
simple (linear) in QED and somewhat less simple (nonlinear) in Yang-Mills
theories, the violent reprocessing of pointlike localized unphysical matter
fields into noncompact string-localized charged fields is a change which the
gauge approach is incapable to describe.

The only formal similarity between the two settings is that the gauge changing
transformations resemble the changes of vectorpotentials under changes of
string directions $e\rightarrow e^{\prime}$ (see end of next section)$.$

In the next section the Wigner theory will be extended in such a way that the
full spectrum of covariant possibilities is realized. This is achieved by
resolving the clash between localization and Hilbert space in favor of
maintaining the latter. This will, with the help of modular localization
theory, automatically result in semiinfinite string-localized
vectorpotentials. Important explicit formulas in terms of string-localized
intertwiners from the unique Wigner representation to covariant string
potentials will be presented. It also will be shown that the use of
vectorpotentials in Stokes theorem applied to magnetic surface fluxes gives
the correct Aharonov-Bohm effect only in case of use of the stringlike
physical vectorpotential.

The last section comments on interactions and indicates how the Higgs
mechanism allows a more physical presentation in terms of the "Schwinger-Higgs
screening" which is a QFT localization analog of the quantum mechanical Debeye
screening in which the Coulomb potential passes to an effective short range
Yukawa potential which in the localization analogy means a pointlike
re-localization of string-localized charge generators via screening. 

The reader who looks for a more detaild account of some of the issues is
referred to \cite{charge}. 

\section{String- localization in Wigner representations from modular
localization}

The representation-theoretical approach to covariance and localization has
been described before \cite{BGL}\cite{MSY}\cite{charge}, therefore the
following presentation will be short and problem-oriented.

Being confused by the many ad hoc linear field equations for relativistic wave
functions which appeared in the 30s, some of them being equivalent to each
other in their physical content, Wigner in 1939 published a completely
intrinsic description in terms of irreducible unitary ray representations of
the Poincar\'{e} group. He found that there are 3 classes of positive energy
representation one massive class with halfinteger spin and two massless
classes, the finite helicity class and infinite helicity class (originally
called "continuous spin" class); the only representation theoretical
difference between the two is the representation theory of its noncompact
little group $E(2)$, the euclidean group in two dimensions which only in the
last case is faithfully represented (its "translations" are nontrivial).

For the first two Wigner and in more detail Weinberg \cite{Wein}, found that
there are many covariant field equations whose covariant representations in
the dotted/undotted ($A,\dot{B}$) representation formalism are connected to
the physical spin/helicity $s$ by%

\begin{align}
\left\vert A-\dot{B}\right\vert  &  \leq s\leq\left\vert A+\dot{B}\right\vert
,\text{ }m>0\label{cov}\\
s  &  =\left\vert A-\dot{B}\right\vert ,~m=0 \label{zero}%
\end{align}
where the second line expresses the fact that, as a result of the nonfaithful
representation of the fix point group $E(2)$ of a lightlike vector, one looses
a large number of covariant generators, including the possibility of
generating the (m=0,s=1) representation by a covariant vectorpotential. The
third infinite spin representation class is the most exotic one; as a result
of a continuous euclidean mass-like parameter $\kappa$ it is, similar to the
massive family, rather big; its physical content was beginning to get
unravelled very recently recently \cite{MSY}. In the two finite spin cases one
can calculate the generating covariant fields in terms of intertwiner
$u^{A\dot{B}}(p)~$which intertwine between the Wigner representation and its
various spinorial representations by group theoretical methods \cite{Wein}$;$
for the infinite spin case there are no pointlike free fields rather all free
fields are string-localized and the intertwiners are infinite component
objects $u(p,e)$ which depend on a string direction $e.$

For zero mass finite spin representations (\ref{zero}) the standard
covariantization only leads to field strength as $F_{\mu\nu}(x)$ for s=1,
$R_{\mu,\nu,\kappa,\lambda}$ for s=2, etc. as the pointlike fields with the
lowest possible dimension $d(s)=s+1.~$The missing $A_{\mu}$-vector and
$g_{\mu\nu}$-tensor potentials do not appear in the covariantization of the
Wigner representation approach. This is a manifestation of a clash between
localization, in this case pointlike localization, and the Hilbert space
setting of QT. The gauge theoretical setting results from this clash by
relaxing the Hilbert space structure and retaining the pointlike localization.
In this case the $A_{\mu}(x)$ lives in an "ghostly" indefinite metric space
(Gupta Bleuler, BRST) and the physical Wigner-Fock space of the field strength
is only recovered by passing to a factor- or cohomology space. In this setting
the vectormeson is a formal device which guaranties the continued use of the
standard pointlike field formalism. The alternative way of resolving the
Hilbert space -- localization clash is to keep the Hilbert space structure and
relax on pointlike localization. It turns out that the only localization which
never leads to problems with the Hilbert space structure is semiinfinite
stringlike i.e. on a halfline: $x+\mathbb{R}_{+}e$ with $e$ being the
spacelike string direction. Allowing this weaker localization one can
construct covariant string-localized vectorpotentials which live in the same
Hilbert space as the pointlike field strength, which are related by
differential operators, and which act cyclically on the vacuum.

Their intertwiners $u(p,e)$ are not easily accessible by group theoretical
methods, here the method of modular localization is more efficient. This
method combines the Wigner representation theoretical setting with that of
modular localization.

First some general remarks about localization. Quantum theory comes with two
notions of localization, the Born-Newton-Wigner localization which corresponds
to a classical \textit{action at a distance} dynamics describing processes
with unlimited basic propagation speed. Its quantum reign is QM\footnote{This
includes relativistic QM \cite{interface} in the form of \textit{direct
particle interactions} (DPI) which is only asymptotically frame-independent
but leading to an invariant S-matrix.} and it comes with a position operator
$\vec{x}_{op}$ whose spectral decomposition leads to wave functions $\psi
(\vec{x})$ and their Born probability density $\left\vert \psi(\vec
{x})\right\vert ^{2}$ Finite velocities as the sound velocity in case of a
system of coupled oscillators are effective velocities showing up in
expectation values in certain states; strictly speaking they become exact in
the asymptotic scattering limit.

This probabilistic localization continues to be present in QFT, but another
more fundamental localization takes over, the modular localization. One may
view it as the quantum counterpart of the causal Cauchy propagation of
classical hyperbolic differential equations. In contradistinction to the BNW
localization which is consistent with QM but not intrinsic to it (Born
introduced it several years after the discovery of QM) the modular
localization is completely intrinsic. This explains why mathematicians
discovered it independently in the form of the Tomita-Takesaki modular theory
of operator algebras whereas physicists came to it through their study of
thermal properties (KMS states) of open systems. The connection with
localization came later and obtained a helping hand from the thermal aspects
of the localization in front of black hole horizons. In interacting QFT there
are no particles in compact spacetime regions, the inexorable presence of
interaction caused vacuum polarization as an epiphenomenon of modular
localization prevents the existence of such localized one particle operators
and states whereas with the frame-dependent BNW localization there are no such
properties. Nevertheless both localizations coalesce in the asymptotic
scattering region, which is crucial for the existence of a
vacuum-polarisation-free frame-independent S-matrix for particles.

Although historically the operator algebraic formulation of modular theory
exited before the (quite recent) modular localization of states, it is easier
and more convenient to present the latter before the former.

The simplest context for a presentation of the idea of modular localization is
in the context of the Wigner representation theory of the Poincar\'{e} group.
It has been realized by Brunetti, Guido and Longo \cite{BGL} and in a more
special context in \cite{AOP} there is a natural localization structure on the
Wigner representation space for any positive energy representation of the
proper Poincar\'{e} group.

The starting point is an irreducible representation $U_{1}~$of the
Poincar\'{e} group including the antiunitary TCP reflection on a Hilbert space
$H_{1}$ that after "second quantization" becomes the single-particle subspace
of the Hilbert space (Wigner-Fock-space) $H_{WF}$ of the associated
interaction free QFT\footnote{The construction works for arbitrary positive
energy representations, not only irreducible ones.}. The construction proceeds
according to the following steps \cite{BGL}\cite{Fa-Sc}\cite{MSY}.

One first fixes a reference wedge region, e.g. $W_{0}=\{x\in\mathbb{R}%
^{d},x^{d-1}>\left\vert x^{0}\right\vert \}$ and considers the one-parametric
L-boost group (the hyperbolic rotation by $\chi$ in the $x^{d-1}-x^{0}$ plane)
which leaves $W_{0}$ invariant; one also needs the reflection $j_{W_{0}}$
across the edge of the wedge which is apart from a $\pi$-rotation in the
transverse plane identical to the TCP transformation. The Wigner
representation $U_{1}$ in $H_{1}$ is then used to define two commuting
wedge-affiliated operators
\begin{equation}
\mathfrak{\Delta}_{1,W_{0}}^{it}=U_{1}(0,\Lambda_{W_{0}}(\chi=-2\pi
t)),~J_{1,W_{0}}=U_{1}(0,j_{W_{0}})
\end{equation}
where attention should be paid to the fact that in a positive energy
representation any operator which inverts time is necessarily
antilinear\footnote{The wedge reflection $\mathfrak{j}_{W_{0}}$ differs from
the TCP operator only by a $\pi$-rotation around the W$_{0}$ axis.}. A unitary
one- parametric strongly continuous subgroup as $\Delta_{1,W_{0}}^{it}$ acting
on $H_{1}$ can be written in terms of a selfadjoint generator as
$\Delta_{1,W_{0}}^{it}=e^{-itK_{W_{0}}}$ and therefore permits an "analytic
continuation" in $t$ to an unbounded densely defined positive operators
$\Delta_{1,W_{0}}^{s}$ with dense domains which decrease with increasing $s$.
Poincar\'{e} covariance allows to extend these definitions to wedges $W$ in
general position, and intersections of wedges lead to the definitions for
general localization regions (see later). Since the localization is clear from
the context, a generic notation without subscripts will cause no confusion.
With the help of this operator one defines the unbounded antilinear operator
$S_{1}$ which has the same dense domain.%
\begin{align}
&  S_{1}=\Delta_{1}^{\frac{1}{2}}J_{1},~domS_{1}=dom\Delta_{1}^{\frac{1}{2}}\\
&  J_{1}\Delta_{1}^{\frac{1}{2}}J_{1}\mathfrak{=}\Delta_{1}^{-\frac{1}{2}}%
\end{align}

Whereas the unitary operator $\Delta_{1}^{it}$ commutes with the reflection,
the antiunitarity of the reflection $J_{1}$ causes a change of sign in the
analytic continuation as written in the second line. This leads to the
involutivity of the $S$-operator as well as the identity of its range with its
domain
\begin{align*}
S_{1}^{2}  &  \subset\mathbf{1}\\
dom~S_{1}  &  =ran~S_{1}%
\end{align*}

Idempotent means that the $S$-operator has $\pm1$ eigenspaces; since it is
antilinear the +space multiplied with $i$ changes the sign and becomes the -
space; hence it suffices to introduce a notation for just one of the two
eigenspaces%
\begin{align}
K_{1}(W) &  =\{domain~of~\Delta_{1,W}^{\frac{1}{2}},~S_{1,W}\psi
=\psi\}\label{K}\\
J_{1W}K_{1}(W) &  =K(W^{\prime})=K_{1}(W)^{\prime},\text{ }duality\nonumber\\
\overline{K_{1}(W)+iK_{1}(W)} &  =H_{1},\text{ }K_{1}(W)\cap iK_{1}%
(W)=0\nonumber
\end{align}

It is important to be aware that, unlike QM, we are dealing here with real
(closed) subspaces $K$ of the complex one-particle Wigner representation space
$H_{1}$.

An alternative which avoids the use of real subspaces is to directly work with
complex dense subspaces as in the third line. Introducing the graph norm of
the dense space, the complex subspace in the third line becomes a Hilbert
space in its own right. The upper dash on regions in the second line denotes
the causal disjoint (which is the opposite wedge) whereas the dash on real
subspaces means the simplectic complement with respect to the simplectic form
$Im(\cdot,\cdot)$ on $H_{1}.$

The two equations in the third line are the defining property of what is
called the \textit{standardness} of a subspace\footnote{According to the
Reeh-Schlieder theorem a local algebra $\mathcal{A(O})$ in QFT is in standard
position with respect to the vacuum i.e. it acts on the vacuum in a cyclic and
separating manner. The spatial standardness, which follows directly from
Wigner representation theory, is just the one-particle projection of the
Reeh-Schlieder property.}; any standard K-space permits to define an abstract
s-operator as follows%
\begin{align}
S_{1}(\psi+i\varphi)  &  =\psi-i\varphi\\
S_{1}  &  =J_{1}\Delta_{1}^{\frac{1}{2}}\nonumber
\end{align}
whose polar decomposition (written in the second line) returns the two modular
objects $\Delta_{1}^{it}$ and $J_{1}$ which outside the context of the
Poincar\'{e} group has in general no geometric significance. The domain of the
Tomita $S$-operator is the same as the domain of $\Delta^{\frac{1}{2}}$ namely
the real sum of the $K$ space and its imaginary multiple. Note that in the
present context this domain is determined solely by Wigner's group
representation theory without any reference to a (nonexistent) covariant
position operator or an extrinsic probability notion.

It is easy to obtain a \textit{net of K-spaces} and associated $S$ operators
by $U_{1}(a,\Lambda)$-transforming the K-space for the distinguished $W_{0}.$
A bit more tricky is the construction of sharper localized subspaces via
intersections
\begin{equation}
K_{1}(\mathcal{O})\equiv%
%TCIMACRO{\dbigcap \limits_{W\supset\mathcal{O}}}%
%BeginExpansion
{\displaystyle\bigcap\limits_{W\supset\mathcal{O}}}
%EndExpansion
K_{1}(W)
\end{equation}
where $\mathcal{O}$ denotes a causally complete smaller region (noncompact
spacelike cone, compact double cone). Intersection may not be standard, in
fact they may be zero in which case the theory allows localization in $W$ (it
always does) but not in $\mathcal{O}.$ Such a theory is still causal, but has
no pointlike localized generators not local in the sense that its associated
free fields are pointlike. However spacelike cone intersections, whose core is
a semiinfinite spacelike string $x+\mathbb{R}_{+}e$ are always standard
\cite{BGL}, which implies that semiinfinite string generating wave functions
(wave function-valued distributions in $x$ and $e$) exist in every positive
energy representation. For the first two classes the $K_{1}$-space is standard
for arbitrarily small $\mathcal{O}$, but this is definitely not the case for
the infinite helicity family for which the compact localization spaces turn
out to be trivial\footnote{It is quite easy to prove the standardness for
spacelike cone localization (leading to singular stringlike generating fields)
just from the positive energy property which is shared by all three families
\cite{BGL}.}.

It is well known that there are two equivalent ways to get from Wigner
representation spaces to interaction-free operators acting in a Wigner-Fock
Hilbert space. The most intrinsic one is the direct functorial relation of the
real subspaces $K_{1}(\mathcal{O})$ to spacetime-indexed net of operator
algebras $\mathcal{A(O}),~\mathcal{O}\subset\mathbb{R}^{4}$. This functorial
relation is often misleadingly called $2nd$ quantization, ignoring that
(Lagrangian) quantization is an art and not a mathematical functor. This
functorial relation is as unique as the Wigner representation theory; the
large choice of possible coordinatizations of these algebras in terms of
operator-valued distributional covariant generators is related to the physical
spin s by the formulae (\ref{cov},\ref{zero}); but these are only the linear
generators of the operator algebras, in addition there are infinitely many
composite generators which can be expressed in terms of Wick-ordered
polynomials of the linear ones. In view of the analogy with the
coordinatization of geometry these operator-valued distribution generators
constitute coordinatizations of spacetime-indexed nets of operator algebras.

The second way is better known \cite{Wein} since it follows directly from the
covariantization of Wigner wave functions. The result in the standard case is
the well known formula for free fields in terms of the u,v intertwiners
between the Wigner representation and their various covariant ($A,\dot{B}$)
spinorial realizations%
\begin{equation}
\Psi^{(A,\dot{B})}(x)=\frac{1}{\left(  2\pi\right)  ^{\frac{3}{2}}}%
\int(e^{-ipx}\sum_{s_{3}=\pm s}u^{(A,\dot{B})}(p)\cdot a(p)+e^{ipx}\sum
_{s_{3}=\pm s}v^{(A,\dot{B})}(p)\cdot b^{\ast}(p))\frac{d^{3}p}{2\omega}%
\end{equation}
Here a,b are the Wigner annihilation,creation operators which depend in
addition to p on the little Hilbert space which is a representation space of
the little group SU(2) or for m=0 E(2), the dot denotes the inner product in
this Hilbert space. As mentioned $dim\Psi^{(A,\dot{B})}\geq s+1,$for $s\geq1$
i.e. no higher spin matter has interactions within the boundaries of the power
counting renormalizability.

There are explicit formulas for the intertwiners which have been derived by
group theory, a method which is however not sufficient in case of
string-localized covariant representations as needed for the (vector, tensor)
potentials in case of m=0. In that case the analog formulas have been derived
in the setting of modular localization. The result can be written in a similar
form. For $(m=0,s\geq1)$ one finds for the covariant string-localized
potentials with $\left\vert A+\dot{B}\right\vert \geq s\geq\left\vert
A-\dot{B}\right\vert $ and the pointlike solution $s=\left\vert A-\dot
{B}\right\vert $ in terms of field strengths excluded

\begin{align}
&  \Psi^{(A,\dot{B})}(x;e)=\frac{1}{\left(  2\pi\right)  ^{\frac{3}{2}}}%
\int(e^{-ipx}\sum_{s_{3}=\pm s}u^{(A,\dot{B})}(p,s_{3};e)a(p,s_{3})+\\
&
\ \ \ \ \ \ \ \ \ \ \ \ \ \ \ \ \ \ \ \ \ \ \ \ \ \ \ \ \ \ \ \ \ \ \ \ \ \ +e^{ipx}%
\sum_{s_{3}=\pm s}v^{(A,\dot{B})}(p,s_{3};e)b^{\ast}(p,s_{3}))\frac{d^{3}%
p}{2\omega}\nonumber
\end{align}
with explicit formulas for the intertwiners which may be pictured as
$(2A+1)(2B+1)\times2$ $~p,e$-dependent rectangular matrices which act on the
two-component (labeled by helicities) columns of creation/annihilation
operators. The only case which is important for the next section is the
vectorpotential s=1%
\begin{align*}
&  A^{\mu}(x,e)=\frac{1}{\left(  2\pi\right)  ^{\frac{3}{2}}}\int(e^{-ipx}%
\sum_{s_{3}=\pm1}u^{\mu}(p,s_{3};e)a(p,s_{3})+e^{ipx}\sum_{s_{3}=\pm
1}\overline{u^{\mu}(p,s_{3};e)}a^{\ast}(p,s_{3}))\frac{d^{3}p}{2\omega}\\
&  u^{\mu}(p,s_{3};e)_{\pm}=\frac{i}{pe+i\varepsilon}\{(\hat{e}_{\mp
}(p)e)p^{\mu}-(pe)\hat{e}_{\mp}^{\mu}(p)\}
\end{align*}
where the $\hat{e}_{\pm}(p)$ denotes the two p-dependent photon polarization
vectors. The notation is also appropriate for the higher (integer) spins which
also can be expressed in terms of higher tensor powers two polarization
vectors and a higher power $\left(  pe+i\varepsilon\right)  ^{-s}$ of the
string-localization causing factor. Once one arrives at the formulas for the
interwiners one can, without knowing anything about modular localization
theory, check is covariance and locality properties%

\begin{align}
&  U(\Lambda)\Psi^{(A,\dot{B})}(x,e)U^{\ast}(\Lambda)=D^{(A,\dot{B})}%
(\Lambda^{-1})\Psi^{(A,\dot{B})}(\Lambda x,\Lambda e)\\
&  \left[  \Psi^{(A,\dot{B})}(x,e),\Psi^{(A^{\prime},\dot{B}^{\prime}%
)}(x^{\prime},e^{\prime}\right]  _{\pm}=0,~x+\mathbb{R}_{+}e><x^{\prime
}+\mathbb{R}_{+}e^{\prime}\nonumber
\end{align}

As expected, the scaling degree of the potential is $d_{sca}(A^{\mu}(x,e))=1$
i.e. better than that of the field strength. The resulting two-point function
is of the form \cite{JRio}%

\begin{align}
&  \left\langle A_{\mu}(x;e)A_{v}(x^{\prime};e^{\prime})\right\rangle =\int
e^{-ip(x-x^{\prime})}W_{\mu\nu}(p;e,e^{\prime})\frac{d^{3}p}{2p_{0}%
},~\curvearrowright\left[  A_{\mu}(x;e)A_{v}(x^{\prime};e^{\prime})\right]
\label{tp}\\
&  W_{\mu\nu}(p;e,e^{\prime})=-g_{\mu\nu}-\frac{p_{\mu}p_{\nu}(e\cdot
e^{\prime})}{(p\cdot e-i\varepsilon)(p\cdot e^{\prime}+i\varepsilon)}%
+\frac{p_{\mu}e_{\nu}}{(e\cdot p-i\varepsilon)}+\frac{p_{\nu}e_{\mu}^{\prime}%
}{(e^{\prime}\cdot p+i\varepsilon)}\nonumber
\end{align}
The presence of the last 3 terms is crucial for the Hilbert space structure;
without them one would fall back into the indefinite metric and negative
probabilities. Since free potentials and free field strength are always
related by linear differential operators, it is not surprising that the
two-point functions of the potentials can be written as an inverse power in
$pe$ times a tensorial expression in $p$ and $e^{\prime}s.$

Instead of the gauge transformation there is now a rule for the change of the
vectorpotential under $e\rightarrow e^{\prime}$%

\begin{align}
&  A^{\mu}(x,e)\rightarrow A^{\mu}(x,e^{\prime})+\partial^{\mu}\Phi
(x;e,e^{\prime})\label{change}\\
&  \Phi(x,e,e^{\prime})=\int e_{\mu}A^{\mu}(x+te^{\prime},e)dt\nonumber
\end{align}
This law for the change of strings continues to be valid in interacting
theories in which the relation between string-localized potentials and
physical (pointlocal) field strength remains linear, it however suffers
interaction dependent modifications for Yang-Mills interactions which are in a
one to one relation with the pointlike nonlinear field strength; in fact
independence of $e$ is synonymous with pointlike localization just as in the
setting of gauge theory where gauge invariance is synonymous with pointlike
generated local observables.

Already in the absence of interactions the unqualified use of the gauge
formalism can lead to wrong results which are avoided in the Hilbert space
string-localized setting. A famous case is the Aharonov-Bohm\footnote{The A-B
effect is a semiclassical effect of quantum mechanical matter in an external
magnetic potential.} effect or rather its specification in the setting of QFT.
From the two-point function (\ref{tp}) one gets the commutator commutator of
the stringlike potentials and from the latter one finds the commutator of the
pointlike field strengths which are independent of $e.$ We only need the
equal-time restriction of the $H-E$ commutator%
\begin{equation}
\left[  H_{i}(\mathbf{x}),E_{j}(\mathbf{x}^{\prime})\right]  \sim
\varepsilon_{ijk}\partial^{k}\delta(\mathbf{x}-\mathbf{x}^{\prime})
\end{equation}
from which one can compute the commutator of two $\rho$-regularized electric
and magnetic delta function fluxes going through two orthogonal disks $D_{1}$
and $D_{2}~$which intersect in such a way that the boundary of one passes
through the center of the other.

Following \cite{LTR} one looks at a situation of two spatially separated, but
interlocking regions $\mathcal{T}_{1}$ and $\mathcal{T}_{2}$ in which one
represents as the smoothened boundary of two orthogonal unit discs. The delta
function fluxes through the $D_{i}$ are smoothened by convoluting $\star$ with
a smooth function $\rho_{i}(\mathbf{x})$ supported in an $\varepsilon$-ball
$B_{\varepsilon}$; the interlocking $\mathcal{T}_{i}$ are then simply obtained
as $\mathcal{T}_{i}=\partial D_{i}+B_{\varepsilon}~i=1,2.$ One computes the
following objects
\begin{align}
&  [\vec{E}(\vec{g}_{1})\vec{H}(\vec{g}_{2})]=\int\vec{g}_{1}(x)rot\vec{g}%
_{2}(x)d^{3}x=\int\rho_{1}(x)d^{3}x\int\rho_{2}(y)d^{3}y\label{alg}\\
&  =\int\rho_{1}(x)d^{3}x\int\rho_{2}(y)d^{3}y~~\ \vec{g}_{i}=\vec{\Phi}%
_{i}\star\rho,~\vec{\Phi}_{i}(\vec{f})=\int_{D_{i}}\vec{f}d\vec{D}%
_{i}\nonumber
\end{align}
The $\Phi_{i}$ is the functional which describe the flux through $D_{i}$, a
kind of surface delta function. Of one would use the curl relation between the
magnetic field and the unphysical pointlike vectorpotential the application of
the Stokes theorem would lead to an integral over the $\mathcal{T}_{i}$ which
vanishes. On the other hand the the same integral in terms of the
string-localized potential gives the correct nonvanishing A-B expression.

This does not mean that the gauge theoretic setting is wrong, but only that
one has to be careful. The safest strategy would be to pass after the
calculations are finished and before the physical interpretation to the gauge
invariant objects; but the difficulty in achieving this for the non pointlike
generated charged matter is the motivation for writing this paper.

\section{Stringlike potentials in interactions,\ Schwinger-Higgs charge
screening}

There are very different ways to introduce interactions; one completely
intrinsic nonperturbative method which starts with classifying generators of
wedge algebras whose properties are very closely related to crossing and
analytic exchange properties of the bootstrap-formfactor program and works its
way down to intersections by forming intersections which lead to arbitrarily
small double cone algebras and their pointlike field generators. This
construction method is very much at its beginnings and has only been
understood in the case of factorizing models in d=1+1. In those cases where it
works it leads to an existence proof and an explicit construction, an
achievement which no other method has attained since the beginning of QFT.

Its intuitive basis is the insight that the weaker the spacetime localization,
the better the control of the ubiquitous vacuum polarizations in the presence
of interactions. Next to the whole spacetime, in which it is possible to find
all operators, including those which applied to the vacuum create pure
one-particle states (without accompanying vacuum polarization clouds), the
best compromise between field (or operator algebra) localization and particles
in the presence of (any kind of) interactions is the (noncompact) wedge region
$W$. Of pivotal importance for this method is the relation between the modular
objects and the scattering matrix (\cite{charge})

QFT's in d=1+3 can presently only be accessed by perturbing free fields with
polynomial interaction densities. In this case the modular localization method
can only be used for free fields. As explained in the previous section this
only leads to new insights in case of ($m=0,s\geq1$) representations i.e. in
case of gauge theories and higher spin tensor potentials. As also explained
there, the use of the string-localized potentials also solves the problem of
keeping the free short distance dimensions at $d_{sd}=1$ thus preventing their
exclusion for reasons of nonexistence of renormalizable interactions. Unlike
the previous nonperturbative method, perturbative series for correlation
functions are known to always diverge i.e. they cannot be used to secure the
mathematical existence of a model; at best they are asymptotically convergent
for infinitesimally small couplings; properties which are true in every order
are believed to indicate structural characteristics of the model.

Perturbation theory is usually formulated in terms of Lagrangian quantization
or the closely related functional integral representations. But is has been
known for a long time \cite{any spin} that Feynman rules do not need free
fields of the Euler-Lagrange type, any type of free field as they arise
through covariantization from Wigner representations will do to be used in a
scalar interaction density of the causal perturbation (Epstein-Glaser)
setting. In fact the string-localized potentials of the previous section are
definitely not Euler-Lagrange and can not be used in any quantization scheme
be it Lagrangian or functional integral, hence the Euler-Lagrange propery is a
restriction required by those special ways of accessing perturbation theory
but not of perturbation theory per se.

In the previous section we learned that the full covariance spectrum
(\ref{cov}) for zero mass finite helicity representation can be regained by
admitting stringlike fields. The pointlike free \textit{field strength}
\footnote{We use this terminology in a generalized sense; all the pointlike
generators (the only ones considered in \cite{Wein}) are called field strength
(generalizing the $F_{\mu\nu}$) whereas the remaining string-localized
generators are named potentials.} is then connected with the free
\textit{stringlike potentials }by covariant differential operators. Both, the
pointlike field strength and the stringlike potentials do not only create the
same Hilbert space from the vacuum, they also fulfill the Reeh-Schlieder
theorem (popular name: state-operator relation) which in case of
string-localization means that the operators with a localization around an
arbitrary small neighborhood of the $(x,e)$ string applied to the vacuum is
dense in the Hilbert space.

We have presented structural arguments in favor of using stringlike potentials
(rather than pointlike field strength) even in the absence of interactions
when Stokes argument is invoked for surface integrals over magnetic fluxes as
in the QFT Aharonov-Bohm argument. The A-B effect is correctly described with
string-localized potentials whereas pointlike potentials lead to a zero
effect. It is also well-known that the Maxwell charges vanish in the standard
gauge indefinite metric setting pointlike potentials and that the presence of
indefinite metric prevents the validity of the Dirac-Maxwell equations
\cite{LTR}\textbf{\cite{charge}}. This is another call to be careful with
drawing physical conclusion in the gauge setting. The rule of gauge theory is
of course to go first to the gauge-invariant correlations and perform a GNS
reconstruction of the corresponding operators in the canonically associated
Hilbert space. Only after having achieved this one can draw physical conclusions.

There is no problem with the subalgebra of pointlike (strictly observable)
generators. But the Maxwell- or Yang-Mills- charge carrying operators are
never pointlike generated and the gauge setting offers no strategy to
construct them. These problems become particularly pressing if one looks at
article and textbooks \cite{text} on QCD where the technical advantage of the
analytic continuation method of dimensional regularization with respect to
gauge theories is misunderstood as making renormalized pointlike quark fields
objects of physical interests.

Of course the presence of the ghost degrees of freedom renders the gauge
setting renormalizable. However the string-localization in a Hilbert space
does the same: for each m=0 spin $s\geq1$ there exists always a potential of
lowest possible dimension namely $d_{sca}(\Psi^{(\frac{s}{2},\frac{\dot{s}}%
{2})}(x,e))=1$ which is the power-counting prerequisite for constructing
renormalizable interactions. Strictly speaking the renormalizabilty of
pointlike would already stop after $s=0,\frac{1}{2};$ beyond there is only the
alternative of either using the gauge approach (which is only known for
m=0,s=1) or the string-localized potential setting for which there are
renormalizable candidates for any s. The short distance dimension of pointlike
objects increase with s; this is well known for the massive case where the
covariance for pointlike fields covers the whole spinorial spectrum
(\ref{cov}) and there is no need for string-localization coming from
representation theory. The simplest example would be a massive pointlike
vector field $A_{\mu}(x)$ with $d_{sd}=2$ whereas the dimension of $A_{\mu
}(x,\mu)$ is $d_{sd}=1.$ It is only the stringlike massive potential which has
a massless limit.

It was already mentioned that the string-localization has hardly any physical
consequences for photons, since even in the presence of interactions the
content of the calculated theory can be fully described in terms of linearly
related pointlike field strengths. Even the scattering theory of photons in
the charge zero sector has no infrared problems and follows a similar logic as
LSZ \cite{phscat}. However the interaction-induced string-localization of the
charged field which is transferred from the
vectorpotentials\footnote{Localization of the free fields, in terms of which
the interaction is defined in the perturbative setting, is not individually
preserved in the presence of interactions; the would be charged fields are not
immune against delocalization from interactions with stringlike
vectorpotentials.} is a much more serious matter; it is inexorably connected
with the electric charge, and there is no linear operation nor any other
manipulation which turns the noncompact localization of charged quantum matter
into compact localization; electrically charged operators have no better
generators than string-localized ones. The argument \cite{Haag} based on the
use of the quantum adaptation of Gauss's law shows that the noncompact (at
best stringlike) localization nature of generating Maxwell charge-carrying
fields is not limited to perturbation theory. The stringlike localization is
so strong that even the Lorentz symmetry becomes spontaneously broken in
nontrivial charge sectors.

It is customary to refer to Maxwell or Yang-Mills theories as \textit{local}
gauge theories and to theories involving complex fields (scalar or spinor) and
which have interactions which are invariant under constant phase or in case of
multiplets under SU(N) groups (e.g. old-fashioned meson-nucleon models) as
invariant under global gauge group, But this is a somewhat treacherous
terminology which only refers to superficial formal aspects but ignores the
deeper physical distinction. From the viewpoint of localization it is just the
other way around, namely the superselection theory which leads to standard
inner symmetries is built on compactly localizable charges (the DHR
superselection theory \cite{Haag}) whereas for noncompact string-localized
charges a superselection theory, if it exists at all, has continuously many
superselected sectors and its inner symmetry is unknown. In other words the
beautiful reconstruction of superselection sectors and charge-carrying field
algebras from their observable shadow (Marc Kac: how to hear the shape of a
drum) is presently limited to compactly localizable (pointlike generated) charges.

Apart from photon field strength, fields are not directly observable; nobody
has ever measured a hadronic field.

Its most dramatic observable manifestation occurs in the scattering of charged
particles. As mentioned before, the infrared peculiarities of scattering of
electrically charged particles, first noted by Bloch and Nordsiek, were
observed at the time as the stringlike Dirac-Jordan-Mandelstam formula from
gauge theory (\cite{charge}), but the two observations remained disconnected
\footnote{The DJM formula is outside the perturbative approach and does not
reveal any physical reason why in a pointlike theory (Maxwell, Yang-Mills)
charge has no pointlike generators, the other reason is that whereas LSZ
scattering theory identifies momentum space scattering amplitudes directly
with \ spacetime limits, the same procedure applied to charges fields gives
zero and a direct spacetime procedure for inclusive cross sections is not
known.}. the standard perturbative gauge formalism (which existed in its
non-covariant unrenormalized form since the time of the B-N paper) was not
capable to address the construction of string-localized physical fields. This
is particularly evident in renormalized perturbation theory which initially
seemed to require just an adaptation of scattering theory \cite{YFS}, but
whose long term consequences, namely a radical change of one-particle states
("infraparticles") and the ensuing loss of a tensor-factorizing Wigner-Fock as
well as the spontaneous breaking of Lorentz invariance and a missing
spin-statistics theorem for infraparticles, were much more dramatic.

Up to now the dramatic conceptual change was patched up by acting as if the
theory is under the LSZ umbrella and counteracting the resistance of the
theory against invalid assumptions by manipulating its resistance with
infrared cutoffs (the spirit of "effective" QFT) and looking for infrared
stable quantities which finally allow a removal of the cutoff (the photon
inclusive cross section). But the last step, namely the direct connection of
the inclusive cross section for charged infraparticles with an asymptotic
limit of string-localized spacetime objects \ But the history of particle
physics shows that whereas it is helpful to think up intelligent placeholders
for conceptual problems which one cannot solve for the time-being, it is
detrimental for progress to leave them up to the cows come home.

The spacetime setting in a theory as QFT, for which everything must be reduced
to its localization principles, is much more important than in QM where
stationary momentum space scattering formulations compete with time-dependent
ones. As mentioned before Coulomb scattering in QM can be incorporated into
any formulation of scattering theory by extracting a diverging phase factor
which results from the long range. Noncompact string-localization is a more
violent change from pointlike generated QFT than long- versus short range
quantum mechanical interactions.

Perturbative scattering (on-shell) processes represented by graphs which do
not contain inner photon lines turn out to be independent of the string
direction $e$ i.e. they appear as if they would come from a pointlike
interaction\footnote{The time-ordered correlation functions, of which they are
the on-shell restriction, are however string-dependent.}. This includes the
lowest order M\o ller- and Bhaba scattering. The mechanism consists in the
application of the momentum space field equation to the $u,v$ spinor wave
functions so that from (\ref{tp}) only the $g_{\mu\nu}$ term in the photon
propagator survives. The terms involving photon lines attached to external
charge lines do however depend on the string directions; these are the same
graphs which in the old infrared investigations were responsible for the
on-shell infrared divergences i.e. e-dependence and the graphical positioning
of infrared behavior is synonymous; the on-shell infrared divergence and the
distributional e-dependence of correlation functions which prohibits to put to
e's on top of each other (short distance limit in the de Sitter space of the
spacelike e-directions) the are two sides of the same coin. In fact the
smearing in e and the careful constructions of "de Sitter composites" are the
additional handles which the string formalism offers. This should give an
interesting powerful separation of ultraviolet and infrared behavior which in
particular in the standard gauge approach to QCD models has been a stumbling
block. None of the gauge theoretic formalisms, not even the much celebrated
dimensional regularization is capable to achieve this; the separation is only
possible in the new string-localized setting.

In the sequel some remarks on the perturbative use of stringlike
vectorpotentials for scalar QED are presented which is formally defined in
terms of the interaction density\footnote{The integral over the interaction
density is formally e-independent.}%
\begin{equation}
g\varphi(x)^{\ast}(\partial_{\mu}\varphi(x))A^{\mu}(x,e)-g(\partial_{\mu
}\varphi(x)^{\ast})\varphi(x)A^{\mu}(x,e) \label{coup}%
\end{equation}
It is also the simplest interaction which permits to explain the Higgs
mechanism as a QED charge-screening. The use of string-localized
vectorpotentials as compared to the standard gauge formalism deflects the
formal problems of extracting quantum data from an unphysical indefinite
metric setting to the ambitious problem of extending perturbation theory to
the realm of string-localized fields. This is not the place to enter a
presentation of (yet incomplete) results of a string-extended Epstein-Glaser
approach. Fortunately this is not necessary if one only wants to raise
awareness about some differences to the standard gauge approach.

It has been known for a long time that the lowest nontrivial order for the
Kallen-Lehmann spectral function can be calculated without the full
renormalization technology of defining time-ordered functions. With the field
equation
\begin{equation}
(\partial^{\mu}\partial_{\mu}+m^{2})\varphi(x)=gA_{\mu}(x,e)\partial^{\mu
}\varphi(x)
\end{equation}
the two-point function of the right hand side in lowest order is of the form
of a product of two Wightman-functions namely the point-localized
$\left\langle \varphi(x)\varphi^{\ast}(y)\right\rangle =i\Delta^{(+)}(x-y)$
and that of the string-localized vectorpotential (\ref{tp})%
\begin{equation}
\left\langle A^{\mu}(x,e)A^{\nu}(x^{\prime},e^{\prime})\right\rangle
\left\langle \partial_{\mu}\varphi(x)\partial_{\nu}\varphi^{\ast}(x^{\prime
})\right\rangle
\end{equation}
leading to the two-point function in lowest (second) order
\begin{equation}
(\partial_{x}^{2}+m^{2})(\partial_{x^{\prime}}^{2}+m^{2})\left\langle
\varphi(x)\varphi^{\ast}(x^{\prime})\right\rangle _{e,e^{\prime}}^{(2)}\sim
g^{2}\left\langle A^{\mu}(x,e)A^{\nu}(x^{\prime},e^{\prime})\right\rangle
\left\langle \partial_{\mu}\varphi(x)\partial_{\nu}\varphi^{\ast}(x^{\prime
})\right\rangle
\end{equation}
which is manifestly $e$-dependent in a way which cannot be removed by linear
operations as in passing from potentials to field strength. One can simplify
the $e$ dependence by choosing collinear strings $e=e^{\prime},$ but the
vectorpotential propagator develops an infrared singularity and in general
such coincidence limits (composites in d=2+1 de Sitter space) have to be
handled with care (although these objects are always distributions in the
string direction i.e. can be smeared with localizing testfunctions in de
Sitter space); just as the problem of defining interacting composites of
pointlike fields through coincidence limits. The infrared divergence can be
studied in momentum space; a more precise method uses the mathematics of wave
front sets\footnote{Technical details as renormalization, which are necessary
to explore these unexplored regions, will be deferred to seperate work.}. This
simple perturbative argument works for the second order two-point function,
the higher orders cannot be expressed in terms of products of Wightman
function but require time ordering and the Epstein-Glaser iteration.

Not all functions of the matter field $\varphi$ are $e$-dependent; charge
neutral composites, as e.g. normal products $N(\varphi\varphi^{\ast}$)$(x)$ or
the charge density are $e$-independent. On a formal level this can be seen
from the graphical representation since a change of the string direction
$e\rightarrow e^{\prime}$ (\ref{change}) corresponds to an abelian gauge
transformation. The divergence form of the change of localization directions
together with the current-vectorpotential form of the interaction reduces the
$e$-dependence of graphs to \textit{vectorpotentials propagators attached to
external charged} lines while all $e$-dependence in loops cancels by partial
integration and current conservation. This is in complete analogy to the
standard statement that the violation of gauge invariance and the cause of
on-shell infrared divergencies on charged lines result from precisely those
external charge graphs; external string-localized vectorpotential lines cause
no problems since they loose their $e$-dependence upon \ differentiation. A
\textit{neutral external composite} as $\varphi\varphi^{\ast}$ on the other
hand does not generate an external charge line; again the gauge invariance
argument parallels the statement that such an external vertex does not
contribute to the string-localization.

Hence both the gauge invariance in the pointlike indefinite metric formulation
and the $e$-independence in the string-like potential formulation both lead to
pointlike localized subtheories\footnote{Note however that the spacetime
interpretation of the $e$ is not imposed. The proponents of the axial gauge
could have seen in in the free two-pointfunction of vectorpotentials and in
all charge correlators if they would have looked at the commutators inside
their perturbative correlation functions. The axial "gauge" is not a gauge in
the usual understanding of this terminology.}. But whereas the embedding
theory (Gupta-Bleuler, BRST) in the first case is unphysical, the string-like
approach uses Hilbert space formulations throughout. The pointlike
localization in an indefinite metric description is a fake. Its technical
advantage is that pointlike interactions, whether in Hilbert space or in a
indefinite metric setting, are treatable with the same well known formalism.
The gauge invariant correlation define (via the GNS construction) a new
Hilbert space which coalesces with the subspace obtained by application of the
pointlike generated subalgebra of the physical string-like formulation to the vacuum.

But whereas the noncompact localized charge-carrying fields are objects of a
physical theory, it has not been possible to construct physical charged
operators through Gupta-Bleuler formalism or BRST cohomological descent. The
difficulty here is that one has to construct non-local invariants under the
nonlinear but formally local acting BRST symmetry. So the simplicity of the
gauge formalism has a high prize when it comes to the construction of
genuinely nonlocal objects as charged fields.

This leaves the globally charge neutral \textit{bilocals} in the visor. Their
description is expected to be given in terms of formal \textit{bilocals which
have a stringlike "gauge bridge"} linking the end points of the formal
bilocals. In contrast to the string-localized single operators it is difficult
to construct them in perturbation theory starting from string-localized free
fields; they are not part of the interacting free fields and they are too far
removed from the form of the interaction. In order to understand the relation
between such neutral bilocals and infraparticles one should notice that in
order to approximate a scattering situations, the "gauge-bridge" bilocals will
have to be taken to the limiting situation of an infinite separation distance,
so that the problem of the infinite stringlike localization cannot be avoided,
since it returns in the scattering situation. The only new aspect of the
proposed approach based on string-localized potentials which requires
attention is that the dependence on the individual string directions $e$ is
distributional i.e. must be controlled by (de Sitter) test function smearing
and moreover that composite limits for coalescing $e^{\prime}s$ can be
defined. But a separation of ultraviolet behavior in the Minkowski space $x$
from the infrared aspects encoded in to the ultraviolet aspects of the de
Sitter $e^{\prime}s$ is just what was missing in the standard approach.

Finally there is the problem of Schwinger-Higgs mechanism in terms of string
localization. The standard recipe starts from scalar QED which has 3
parameters (mass of charged field, electromagnetic coupling and quadrilinear
selfcoupling required by renormalization theory). The QED model is then
modified by Schwinger-Higgs screening in such a way that the Maxwell structure
remains and the total number of degrees of freedom are preserved. The standard
way to do this is to introduce an additional parameter via the
vacuumexpectation value of the alias charged field and allow only
manipulations which do not alter the degrees of freedom. We follow Steinmann
\cite{Stei}, who finds that the screened version consists of a selfcoupled
real field $R$ of mass $M$ coupled to a vectormeson $A^{\mu}$ of mass $m$ with
the following interaction%
\begin{align}
L_{int} &  =gmA^{\mu}A_{\mu}R-\frac{gM^{2}}{2m}R^{3}+\frac{1}{2}g^{2}A_{\mu
}A^{\mu}R^{2}-\frac{g^{2}M^{2}}{8m^{2}}R^{4}\label{scre}\\
\Psi &  =R+\frac{g}{2m}R^{2}%
\end{align}
The formula in the second line is obtained by applying the prescription
$\varphi\rightarrow\left\langle \varphi\right\rangle +R+iI$ to the complex
field within the neutral (and therefore point-local) composite $\varphi
\varphi^{\ast}$ and subsequently formally eliminating the $I$ field by a gauge
transformation. The result is the above interaction where $A_{\mu}$ and $R$
are now massive fields. Since the field $\Psi$ is the image of a pointlike
$\varphi\varphi^{\ast}$ under the Higgs prescription, the real matter field
$\Psi,$ as the screened version of the pointlocal charge-neutral
$\varphi\varphi^{\ast}$remains local. However the screening does not only
break the charge symmetry (thus trivializing the charge) and uses the $I$
degrees of freedom to convert the photon into a massive vectormeson, but also
disrupts the even-odd symmetry $R\rightarrow-R$ of the remaining
$R$-interaction. It is the absence of this $\mathbb{Z}_{2}$ selection rule
which transfers the pointlike localization of $\Psi$ to $R$ so that together
with the pointlike $F_{\mu\nu}$\footnote{From the pointlike $F_{\mu\nu}$ on
can construct a pointlike $A_{\mu}(x)$ with the same dimension.} from the
stringlike $A_{\mu}(x,e)$ the screened model is generated in terms of only
pointlike fields. 

Hence in the present context the string-localized potentials, as well as the
gauge theoretic BRST formalism, behave as a "catalyzer" which makes a theory
amenable to renormalization. The former have the additional advantage over in
the latter that the Hilbert space is present throughout the calculation.

One has to be careful in order not to confuse computational recipes with
physical concepts. Nonvanishing vacuum expectations (one-point functions) are
part of a recipe and \textit{should not be directly physically interpreted},
rather one should look at the intrinsic observable consequences\footnote{These
are properties which can be recovered from the observables of the model i.e.
they do not depend on the particular prescription use in its construction.}
before doing the physical mooring. The same vacuum expectation trick applied
to the Goldstone model of spontaneous symmetry breaking has totally different
consequences from its application in the Higgs-Kibble (Brout-Englert,
Guralnik-Hagen) symmetry breaking.

In the case of spontaneous symmetry breaking (Goldstone), the \textit{charge
associated with the conserved current diverges} as a result of the presence of
a zero mass Boson which couples to this current. On the other hand in the
Schwinger-Higgs screening situation \textit{the charge of the conserved
current vanishes} (i.e. is completely screened) and hence there are no charged
objects which would have to obey a charge symmetry with the result that the
lack of charge resulting from a screened Maxwell charge looks like a symmetry breaking.%

\begin{align*}
&  Q_{R,\Delta R}=\int d^{3}xj_{0}(x)f_{R,\Delta R}(x),~~f_{R,\Delta R}(x)=%
\begin{array}
[c]{c}%
1~for~\left\vert \mathbf{x}\right\vert <R\\
0~~for~\left\vert \mathbf{x}\right\vert \geq R+\Delta R
\end{array}
\\
&  \lim_{R\rightarrow\infty}Q_{R,\Delta R}^{spon}\left\vert 0\right\rangle
=\infty,~m_{Goldst}=0;~~\lim_{R\rightarrow\infty}Q_{R,\Delta R}^{screen}%
\psi=0,\text{ }all~m>0
\end{align*}

That the recipe for both uses a shift in field space by a constant does not
mean that the physical content is related. The result of screening is the
vanishing of a Maxwell charge which (as a result of the charge superselection)
allows a copious production of the remaining $R$-matter. 

Successful recipes are often placeholders for problems whose better
understanding needs additional conceptual considerations. In both cases one
can easily see that the incriminated one-point vacuum expectation has no
intrinsic physical meaning, i.e. there is nothing in the intrinsic properties
of the observables of the two theories which reveals that a nonvanishing
one-point function was used in the recipe for its construction. For a detailed
discussion of these issues see \cite{Swieca}.

The premature interpretation in terms of objects which appear in calculational
recipes tends to lead to mystifications in particle theory; in the present
context the screened charged particle has been called the "God particle". As
mentioned before the Schwinger-Higgs screening is analog to the quantum
mechanical Debeye screening in which the elementary Coulomb interaction passes
to the screened large distance effective interaction which has the form of a
short range Yukawa potential. The Schwinger-Higgs screening does not work
(against the original idea of Schwinger) directly with spinor- instead of
scalar matter. If one enriches the above model by starting from QED which
contains in addition to the charged scalar fields also charged Dirac spinors
then the screening mechanism takes place as above via the scalar field which
leads to a loss (screening, bleaching) of the Maxwell charge while the usual
charge superselection property of complex Dirac fields remains unaffected.

The Schwinger-Higgs mechanism has also a scalar field multiplet generalization
to Yang-Mills models; in this case the resulting multicomponent pointlike
localized massive model is much easier to comprehend than its "charged"
string-localized origin. As the result of screening there is no unsolved
confinement/invisibility problem resulting from nonabelian string-localization.

The Schwinger Higgs screening suggests an important general idea about
renormalizable interactions involving massive $s\geq1$ fields, namely that
formal power-counting renormalizability ($d_{sd}=1$) is not enough. For
example a pure Yang-Mills interaction with massive gluons (without an
accompanying massive real scalar multiplet) could be an incorrect idea because
the string-localization of the Hilbert space compatible gluons could spread
all over spacetime or there may exist other reasons why the suspicion that
such theories are not viable may be correct. Such a situation would than be
taken as an indication that a higher spin massive theory would always need
associated lower spin massive particles in order to be localizable; in the s=1
case this would be the s=0 particle resulting from Schwinger-Higgs screening.
Before one tries to understand such a structural mechanism which requires the
presence of localizing lower spin particles it would be interesting to see
whether these new ideas allow any renormalizable s=$\frac{3}{2}$
(Rarita-Schwinger) theories. Even though there may be many formal
power-counting renormalizable massive $s\geq1$ interactions only a few are
expected to be pointlike localized.

It is interesting to mention some mathematical theorems which support the
connection between localization and mass spectrum. The support for placing
more emphasis on localization in trying to conquer the unknown corners of the
standard model comes also from mathematical physics. According to Swieca's
theorem \cite{screening}\cite{Swieca} one expects that the screened
realization of the Maxwellian structure is local i.e. the process of screening
is one of reverting from the electromagnetic string-localization back to point
locality together with passing from a gap-less situation to one with a mass
gap. Last not least the charge screening leads to a Maxwell current with a
vanishing charge\footnote{Swieca does not directly argue in terms of
localization but rather uses the closely related analyticity properties of
formfactors.} and the ensuing copious production of alias charged particles.
The loss of the charge superselection rule in the above formulas (\ref{scre})
is quite extreme, in fact even the $R\leftrightarrow-R$ selection rule has
been broken (\ref{scre}) in the above Schwinger-Higgs screening phase
associated with scalar QED. The general idea for constructing renormalizable
couplings of massive higher spin potentials interacting with themselves or
with normal s=0,1/2 matter cannot rely on a Schwinger-Higgs screening picture
because without having a pointlike charge neutral subalgebra for zero mass
potentials as in QED, which is the starting point of gauge theory, there is no
screening metaphor which could preselect those couplings which have a chance
of leading to a fully pointlike localized theory, even though
renormalizability demands to treat all $s\geq1$ as stringlike objects with
$d_{sca}$=1. Of course at the end of the day one has to be able to find the
renormalizabe models which maintain locality of observables either in the zero
mass setting as (charge-neutral) subalgebras (QED,Yang-Mills) or the massive
theories obtained from the former with the help of the screening idea. gauge
theory is a crutch whose magic power is limited to s=1, for s$>1$ it lost its
power and one has to approach the localization problem directly.

The existence of a gauge theory counterpart, namely the generalization of the
BRST indefinite metric formalism to higher spins, is unknown. So it seems that
with higher spin one is running out of tricks, hence one cannot avoid confront
the localization problem of separating theories involving string-localized
potentials which have pointlike generated subalgebras from those which are
totally nonlocal and therefore unphysical. This opens a new chapter in
renormalization theory and its presentation would, even with more results than
are presently available, go much beyond what was intended under the modest
title of this paper.

An understanding of the Schwinger-Higgs screening prescription in terms of
localization properties should also eliminate a very unpleasant previously
mentioned problem which forces one to pass in a nonrigorous way between the
renormalizable gauge (were the perturbative computations take place) and the
"unitary gauge" which is used for the physical interpretation. The relation
between the two remains somewhat metaphoric.

The screened interaction between a string-localized massive vectorpotential
and a real field (\ref{scre}) remains pointlike because the string
localization of the massive vectorpotential only serves to get below the power
counting limit but does not de-localize the real matter field; since the
pointlike field strength together with the real scalar field generate the
theory, the local generating property holds. In an approach based on
string-localization there is only one description which achieves its
renormalizability by string-localized potentials.

The BRST technology is highly developed, as a glance into the present
literature \cite{Du} shows. It certainly has its merits to work with a
renormalization formalism which starts directly with massive vectormesons
\cite{Du-S} instead of the metaphoric "photon fattened on the Higgs one point
function". It is hard to think how the BRST technology for the presentation of
the Schwinger-Higgs screening model which starts with a massive vectormeson in
\cite{Du} can be improved. For appreciating this work it is however not
necessary to elevate \ "quantum gauge symmetry" (which is used as a technical
trick to make the Schwinger-Higgs mechanism compliant with renormalizability
of massive s=1 fields) from a useful technical tool to the level of a new principle.

Besides the Schwinger-Higgs screening mechanism which leads to renormalizable
interactions of massive vectormesons with low spin matter, there is also the
possibility of renormalizable "massive QED" which in the old days \cite{Lo-S}
was treated within a (indefinite metric) gauge setting in order to lower the
short distance dimension of a massive vectormeson from $d_{sd}=2$ to $1$, and
in this way stay below the power-counting limit. Such a construction only
works in the abelian case; for nonabelian interactions the only way to
describe interacting massive vectormesons coupled to other massive s=0,1/2
quantum matter is via Higgs scalars in their Schwinger-Higgs screening role.
Whereas the local Maxwell charge is screened, the global charges of the
non-Higgs complex matter fields are preserved. It seems that Schwinger's
original idea of a screened phase of spinor QED cannot be realized, at least
not outside the two-dimensional Schwinger model (two-dimensional massless QED).

But the educated conjectures in this section should not create the impression
that the role of the Schwinger-Higgs screening in the renormalizability of
interactions involving selfcoupled massive vectormesons has been completely
clarified; if anything positive has been achieved, it is the de-mystification
of the metaphor of a spontaneous symmetry breaking through the
vacuumexpectation of a complex gauge dependent field and the tale of "God's
particle" which creates the masses of s=1/2 quantum matter. Actually part of
this de-mystification has already been achieved in \cite{Du}.

This leads to the interesting question whether, apart from the presence of the
Higgs particle (the real field as the remnant of the Schwinger-Higgs
screening), there could be an intrinsic difference in the structure of the
vectormeson. Such a difference could come from the fact that the screening
mechanism does not destroy the algebraic structure of the Maxwell equation,
whereas an interaction involving a massive vectormeson coming in the indicated
way from a S-H screening mechanism and interacting with spinorial matter
fields maintains the Maxwell structure. In the nonabelian case this problem
does not arise since apparently the Schwinger-Higgs screening mechanism is the
only way to reconcile renormalizability with localizability (or a return to
physics from an indefinite metric setting).

This raises the interesting question whether renormalizability and pointlike
locality of interactions with massive higher spin $s>1$ potentials is always
related to an associated zero mass problem via an analog of a screening
mechanism in which a lower spin field plays the analog of the Higgs field?

Whereas for interactions between spin one and lower spin fields the physical
mechanism behind the delocalization of matter (or rather its noncompact
re-localization) is to some degree understood, this is not the case for
interacting higher spin matter. Stringlike interactions enlarge the chance of
potentially renormalizable (passing the power counting test) theories, in fact
stringlike potentials with dimension $d_{sca}=1$ exist for any spin (hence
infinitely many) whereas the borderline for pointlike interaction is $s=1/2$
and with the help of the gauge setting $s=1$. Certain interactions, as the
Einstein-Hilbert equation of classical gravity probably remain outside the
power-counting limit even in the stringlike potential setting, but certain
polynomial selfinteractions between the $g_{\mu\nu}(x,e)$ with $dimg_{\mu\nu
}(x,e)=1$ may be renormalizable. The existence of free pointlike field
strength (in this case the linearized Riemann tensor) indicates that there may
be renormalizable interactions which lead to pointlike subalgebras, but the
presence of self-couplings modifies the transformation law under a change of
$e$ (\ref{change}) which now depends on the interaction as it is well-known
from the gauge theoretical formulation for Yang-Mills couplings.

One of course does not know whether QFT is capable to describe quantum gravity
(it never has been tried), but if it does in a manner which is compatible with
renormalized perturbation theory, there will be no way to avoid
string-localized tensorpotentials even if the theory contains linear or
nonlinear related pointlike localized field strength. The trick of gauge
theory, by which one can extract pointlike localized generators without being
required to construct first the string-localized ones, is a resource which
does not seem to exist for higher spins, not even if one is willing to cope
with unphysical ghosts in intermediate steps. The most interesting
interactions are of course the selfinteractions between $(m=0,s>1).$ Here one
runs into similar problems as with Yang-Mills models (next section). The
independence on $e^{\prime}s$ of the local observables leads to nonlinear
transformation laws which extend that of free stringlike potentials and the
non-existence of linear local observables. Although saying this does not solve
any such problem, the lack of an extension of the gauge idea to higher spin
makes one at least appreciative of a new view based on localization.

There is one important case which we have left out, namely that of massless
Yang-Mills theories interaction with massive matter. This will be commented on
in the next section.

There are 2 different categories of delocalization: string-localization with
nontrivial pointlike-generated subalgebras as QED. But generically the
coupling of string-localized fields leads to a theory with \textit{no local
observables}. The models of physical interest are those which contain
nontrivial $e$-independent subfields. For the case at hand the crucial
relation is that the change in the string direction can be written as a
derivative as in (\ref{change}). Interactions which are not invariant under
law of change of $e$ do not give rise to compact localized observables and
physically uninteresting, even if they mathematically exist.

\section{Concluding remarks}

The guiding model for the presentation of the modular localization alternative
to the standard gauge setting has been QED. It would be very interesting to
understand the physical consequences of the much stronger string-localization
in QCD. The third Wigner class of infinite spin representations is a
kinematical illustration of a strong string-localization. Such a string
remains invisible to any local or quasilocal registering counter. The reason
is that such a counter is limited to register the presence of a local piece of
the string but the caused local change would be in contradiction with the
holistic nature of an irreducible object. Such objects cannot be created from
pointlike or QED charge strings. According to popular argument (without
mathematical support) an interaction cannot change the irreducible particle
components, if in the interaction density there was no infinite spin component
there will be none in the resulting theory.

Whereas for gauge theories the use of string-localized potentials can be
considered as a refinement of gauge theory, the use of s=2 string-localized
potentials $g_{\mu\nu}(x,e)$ with $d_{sc}=1$ which, without interactions, is
linearly related to 4-tensor $R_{\mu\nu\kappa\lambda}$ with the symmetry
properties of the Riemann tensor is virgin soil, since in this case there is
no gauge setting. Another more speculative idea envisages a of obtaining
renormalizable higher massive theories by starting from zero mass and
generalizing the screening idea. Since one cannot expect that generic
couplings of higher spin massive string-localized potentials lead to pointlike
generated subalgebras (not even for s=1), one perhaps needs compensating lower
spin fields (in analogy to the screened charged field) in order to select such
theories.  

Leaving the issue of confinement/invisibility aside, one can certainly study
perturbative QCD in the new setting. Since now localization is the central
issue, the only compatible method is the Epstein-Glaser iteration. This
involves a new problem namely the causal ordering (time-ordering) for strings.
For coplanar strings, which are orthogonal to a timelike vector, this can be
achieved \cite{JRio}. Without having done detailed calculation it is not clear
whether this lack of covariance is a blessing (the spontaneously broken
Lorentz invariance in the charge sectors) or a curse.

For the perturbation theory itself there are two possible strategies. On could
either stay close to the spirit of the axial gauge approach;iIn that case one
would try to take the composite limit $e_{i_{1}}....e_{i_{k}}\rightarrow
e_{i}$ so that the $i^{th}$ charge line has just one string direction $e_{i}$
which can then be used to control the infrared properties. On the other hand
there is the setting of Bogoliubov's S functional which would favor an
approach in which the points $e$ on de Sitter space receive the same treatment
the $x^{\prime}s$ in Minkowski space. In that case one would integrate over
all $e^{\prime}s$ which would seem to totally delocalize the strings. Of
course on could try to reintroduce the string-localization by working with
string-localized matter fields and integrate, as one does with the $x,$ over
all internal $e$ of a graph. The fact tha such a procedure does not speak
against it, as long as the S-matrix stays the same. More details and hopefully
the resolution of these questions will be contained in a forthcoming
collaboration \cite{B+J}.

\end{document}